\documentclass[prb,onecolumn,epsfig,]{revtex4}
\usepackage{amsfonts}
\usepackage{amsbsy}
\usepackage{amssymb}
\usepackage{graphicx}
\usepackage[breaklinks,colorlinks = true,linkcolor = red,urlcolor=red,citecolor=red]{hyperref}
\usepackage{subfigure}

\begin{document}

\title{Spin-Orbital Locking, Emergent Pseudo-Spin, and Magnetic order in Honeycomb Lattice Iridates.}
\date{\today}
\author{Subhro Bhattacharjee$^{1,2}$, Sung-Sik Lee$^{2,3}$, and Yong Baek Kim$^{1,4}$}

\address{
$^1$ Department of Physics, University of Toronto, Toronto, Ontario, Canada M5S 1A7.\\
$^2$ Department of Physics \& Astronomy, McMaster University, Hamilton, Ontario, Canada  L8S 4M1.\\ 
$^3$ Perimeter Institute for Theoretical Physics, Waterloo, Ontario, Canada N2L 2Y5.\\
$^4$ School of Physics, Korea Institute for Advanced Study, Seoul 130-722, Korea.
}

\begin{abstract}
The nature of the effective spin Hamiltonian and magnetic order in the honeycomb iridates is explored by considering a trigonal crystal field effect and spin-orbit coupling. Starting from a Hubbard model, an effective spin Hamiltonian is derived in terms of an emergent pseudo-spin-1/2 moment in the limit of large trigonal distortions and spin-orbit coupling. The present pseudo-spins arise from a {\em spin-orbital locking} and are different from the $j_{\rm eff}=1/2$ moments that are obtained when the spin-orbit coupling dominates and trigonal distortions are neglected. The resulting spin Hamiltonian is anisotropic and frustrated by further neighbour interactions. Mean field theory suggests a ground state with 4-sublattice {\em zig-zag} magnetic order in a parameter regime that can be relevant to the honeycomb iridate compound Na$_2$IrO$_3$, where similar magnetic ground state has recently been observed. Various properties of the phase, the spin-wave spectrum and experimental consequences are discussed. The present approach contrasts with the recent proposals to understand iridate compounds starting from the strong spin-orbit coupling limit and neglecting non-cubic lattice distortions. 
\end{abstract}

\maketitle

\section{Introduction}
Interplay between strong spin-orbit (SO) coupling and electron-electron interaction in correlated electron  systems has been a recent subject of intensive study \cite{2007_okamoto,2008_kim,2009_kim,2010_singh,2011_liu,2011_singh,2011_zhao, 2011_tafti,2009_shitade,2010_chaloupka,2011_jiang,2009_jin,2009_podolsky,2011_podolsky,2010_pesin,2010_yang,2010_wan,2011_wang,2010_chen,2006_nakatsuji,2008_lawler2,2008_chen,2009_norman,2011_william}.
 In particular, $5d$ transition metal (e.g.~Iridium (Ir) or Osmium (Os)) oxides are regarded as ideal playgrounds for observing such cooperative effects \cite{2007_okamoto,2008_kim,2009_kim,2010_singh,2011_liu,2011_singh,2006_nakatsuji,2011_zhao,2011_tafti,2009_shitade,2010_chaloupka,2011_jiang,2009_jin,2008_lawler2,2008_chen,2009_norman,2009_podolsky,2011_podolsky,2010_pesin,2010_yang,2010_wan,2011_william,2011_wang,2010_chen}. Compared to $3d$ transition metal oxides, the repulsive Coulomb energy scale in these systems is reduced by the much larger extent of $5d$ orbitals, while the SO coupling is enhanced due to high atomic number ($Z=77$ for Ir and $z=76$ for Os). Moreover, owing to the extended $5d$ orbitals, these systems are very sensitive to the crystal fields. As a result, the energy scales mentioned above often become comparable to each other, leading to a variety of competing phases. Precisely for this reason, one expects to see newer emergent quantum phases in such systems. Indeed, there have been several theoretical proposals, in context of concrete experimental examples, for spin liquids \cite{2007_okamoto,2010_chaloupka,2008_lawler2,2011_podolsky,2010_pesin,2011_jiang,2009_norman}, topological insulators {\cite{2009_shitade,2010_pesin,2010_yang}}, Weyl semimetals \cite{2010_wan,2011_william}, novel magnetically ordered Mott insulators {\cite{2008_kim,2009_kim,2010_singh,2011_liu,2011_singh,2011_wang,2010_chen}} and other related phases {\cite{2011_zhao, 2011_tafti,2011_wang}} in Iridium and Osmium oxides.
 
A typical situation in the iridates consist of Ir$^{+4}$  atoms sitting in the octahedral crystal field of a chalcogen, typically oxygen or sulphur \cite{2007_okamoto,2008_kim,2009_kim,2010_singh,2006_nakatsuji}. This octahedral crystal field splits the five $5d$ orbitals of Ir into the doubly degenerate $e_g$ orbitals and the triply degenerate $t_{2g}$ orbitals (each orbital has a further two-fold spin degeneracy). The $e_g$ orbitals are higher in energy with the energy difference being approximately $3\ {\rm eV}$. There are 5 electrons in the outermost $5d$ shell of Ir$^{+4}$ which occupy the low lying $t_{2g}$ orbitals and the low energy physics is effectively described by projecting out the empty $e_g$ orbitals \cite{1999_fazekas}. A characteristic feature of most of the approaches used to understand these compounds is to treat the SO coupling as the strongest interaction at the atomic level; {\em i.e.}, by considering the effect of extremely strong SO coupling for electrons occupying the $t_{2g}$ orbitals. This decides the nature of the participating atomic orbitals in the low energy effective theory. In this limit, the orbital angular momentum, projected to the $t_{2g}$ manifold, carries an effective orbital angular momentum $l_{\rm eff}=1$ with a negative SO coupling constant \cite{2008_kim,2008_chen,2009_kim}. The projected SO coupling splits the $t_{2g}$ manifold into the lower $j_{\rm eff}=3/2$ quadruplet and the upper $j_{\rm eff}=1/2$ doublet. Out of the five valence electrons, four fill up the quadruplet sector leaving the doublet sector half filled. Thus, in the limit of very strong SO coupling, the half filled doublet sector emerge as the correct low energy degrees of freedom. Considering the effect of coulomb repulsion within a Hubbard model description and performing strong coupling expansion, various spin Hamiltonians for $j_{\rm eff}=1/2$ are then derived within a strong-coupling perturbation theory \cite{2010_chaloupka,2010_yang,2011_wang}.

\begin{figure}
\centering
\includegraphics[scale=0.5]{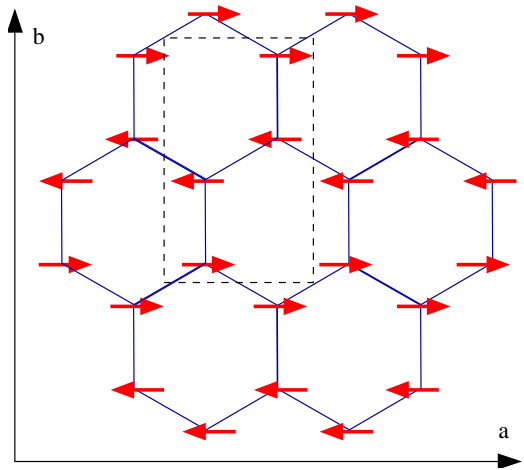}
\label{fig_zig_zag}
\caption{The {\em zig-zag} magnetic structure as found in Ref. \cite{2011_liu}. The magnetic unit cell has 4 sites.}
\end{figure}

In this paper, we, however, consider a different limit where the oxygen octahedra surrounding the Ir$^{+4}$ ions are highly distorted. While the above scenario of half filled $j_{\rm eff}=1/2$ orbitals is applicable to undistorted case, as we shall see, it breaks down in presence of strong distortions of the octahedra. In particular, we consider the effect of trigonal distortions, which may be relevant for some of the iridate systems including the much debated honeycomb lattice iridate, Na$_2$IrO$_3$. We show that, in this limit, a different ``doublet" of orbitals emerge as the low energy degree of freedom. This doublet forms a pseudo-spin-$1/2$ that results from a kind of (physical)spin-orbital {\em locking}, so that the spin and orbital fluctuations are not separable (as discussed below). We emphasize that this pseudo-spin is different from the $j_{\rm eff} = 1/2$ doublet discussed above. The spin Hamiltonian for these pseudo-spins (Eq.~\ref{eq_hamiltonian}), on a honeycomb lattice, admits a 4-sub-lattice {\em zig-zag} (fig. \ref{fig_zig_zag}) pattern in a relevant parameter regime. Such magnetic order has been recently observed in the experiments \cite{2011_liu,2012_choi} on Na$_2$IrO$_3$ and hence our theory may be applicable to this material.

The distortion of the octahedron surrounding the Ir$^{+4}$ generates  a new energy scale associated with the change in the crystal field, which, as we shall see, competes with the SO coupling. Several kinds of distortion may occur, of which we consider the  trigonal distortions of the octahedron where it is stretched/compressed along the body diagonal of the enclosing cube \cite{1999_fazekas}. In the absence SO coupling, such  trigonal crystal field splits the $t_{2g}$ manifold into $e'_{g}$ (with two degenerate orbitals $e'_{1g}$ and $e'_{2g}$) and non-degenerate $a_{1g}$ (again there is an added two-fold spin degeneracy for each of these orbitals). The $e'_{g}$ and $a_{1g}$ levels are respectively occupied by three and two electrons in Ir$^{4+}$. For large trigonal distortions, the splitting between them is big and the $a_{1g}$ orbitals can be projected out. Now, if one adds SO coupling, the low energy degrees of freedom is described by a subspace of the $e'_g$ orbitals which form an emergent pseudo-spin-1/2 doublet out of $|e'_{1g}, \downarrow \rangle$ and $|e'_{2g} \uparrow \rangle$ states, where $\uparrow$ and $\downarrow$ represent the physical spin $s_z=1/2,-1/2$ (the spins are quantized along the axis of trigonal distortion). These pseudo-spin-$1/2$ is different from the $j_{\rm eff}=1/2$ and $j_{\rm eff}=3/2$ multiplets in the strong SO coupling limit as discussed above. Notice the (physical)spin-orbital {\it locking} for the pseudo-spins, as alluded above. 

The two approaches, described in the last two paragraphs, of arriving at the low energy manifold are mutually incompatible. This can be seen as follows: In presence of sizeable trigonal distortions the $j_{\rm eff}=1/2$ and $j_{\rm eff}=3/2$ multiplets mix with each other and can no longer serve as good low energy atomic orbitals. This dichotomy becomes quite evident in the recent studies of Na$_2$IrO$_3$, where, the $Ir^{+4}$ form a honeycomb lattice. Taking into account the strong SO coupling of Ir$^{4+}$ in Na$_2$IrO$_3$, proposed are a model for a topological insulator in the weak interaction limit \cite{2009_shitade} and a Heisenberg-Kitaev (HK) model for a possible spin liquid phase in the strong coupling limit \cite{2010_chaloupka}. These proposals prompted several experimental \cite{2010_singh,2011_liu,2011_singh} and theoretical efforts \cite{2011_jiang, 2011_reuter} to understand the nature of the ground state in this material. Subsequently, it was found that Na$_2$IrO$_3$ orders magnetically at low temperatures \cite{2010_singh}. However, the magnetic moments form a ``zig-zag" pattern (fig. \ref{fig_zig_zag}) which is not consistent with the ones that would be obtained by adding a weak interaction in a topological insulator (canted antiferromagnet) \cite{2009_shitade} or from a nearest neighbour HK model (spin liquid or the so-called stripe antiferromagnet) \cite{2010_chaloupka}. While recent studies {\cite{2011_kimchi}} show that a `zig-zag' order may be stabilized within the HK model by including substantial second and third neighbour antiferromagnetic interactions, it is hard to justify such large further neighbour exchanges without lattice distortions. (An alternate explanation that we do not pursue here is significant charge fluctuations which would mean that the compound is close to metal-insulator transition. The resistivity data seems to support the fact that this compound is a good insulator. \cite{2010_singh}) If lattice distortions are responsible for the significant further neighbour exchanges, then, there would be sizeable distortion of the oxygen octahedra, which in turn may invalidate the above $j_{\rm eff}=1/2$ picture and thus the basic paradigm of the HK model, by mixing the $j_{\rm eff}$=1/2 and $j_{\rm eff}=3/2$ subspaces. Recent finite temperature numerical calculations on the HK model \cite{2011_reuter} also suggest possible inconsistencies with experiments on Na$_2$IrO$_3$.  This necessitates the need for a different starting point, to explain the magnetic properties of Na$_2$IrO$_3$. 

The rest of this paper is organized as follows. In Section \ref{sec_hamiltonian}, we derive the effective spin Hamiltonian in limit of the large trigonal distortion and large SO coupling. This is done by taking the energy scale associated with trigonal distortion to infinity first, followed by that of the SO energy scale. This order of taking the limit gives a spin Hamiltonian in terms of emergent pseudo-spin$-1/2$, which is different from the HK model. This Hamiltonian is both anisotropic and frustrated. It also has further neighbour interactions, the effect of which are enhanced due to anisotropy that makes some of the nearest neighbour bonds weaker. The origin of this anisotropy is trigonal distortion. We argue that this limit may be more applicable for the compound Na$_2$IrO$_3$. Having derived the spin Hamiltonian, we calculate the phase diagram and the spin wave spectrum within mean field theory in Sec.~\ref{sec_meanfield}. We see that the `zig-zag' phase occurs in a relevant parameter regime. We also point out the experimental implications of our calculations in context of Na$_2$IrO$_3$. Finally we summarize the results in Sec.~\ref{sec_summary}. The details of various calculations are given in various appendices.

\section{The Effective Hamiltonian}
\label{sec_hamiltonian}
In the cubic environment the $t_{2g}$ orbitals are degenerate when there is no SO coupling.  Trigonal distortion due to compression or expansion along one of the four $C_3$ axes of IrO$_6$ octahedra lifts this degeneracy. Although it is possible that the axes of trigonal distortions are different in different octahedra \cite{2010_yang}, we find that uniform distortions are more consistent with the experiments (see below) on Na$_2$IrO$_3$.  Hence we consider uniform trigonal distortion.

\subsection{The Trigonal Hamiltonian}
Let us denote the axis of this uniform trigonal distortion of the octahedron by the unit vector $\hat n=\frac{1}{\sqrt{3}}[n_1,n_2,n_3]$, where $n_\alpha=\pm 1$. Since there are 2 directions to each of the 4 trigonal axes we may choose a ``gauge" to specify $\hat n$. This is done by taking $n_1 n_2 n_3=+1$. The Hamiltonian for trigonal distortion, when projected in the $t_{2g}$ sector, gives {\cite{2010_yang}} (in our chosen {\em gauge})
\begin{eqnarray}
H^{t_{2g}}_{tri}=-\sum_i \frac{\Delta_{tri}}{3}\Psi_i^\dagger\left[\begin{array}{ccc}
0 & n_3 & n_2\\
n_3 & 0 & n_1\\
n_2 & n_1 & 0\\
\end{array}\right]\Psi_i,
\label{eq_tri_ham}
\end{eqnarray}
where $\Psi_i^\dagger=[d^\dagger_{yz},d^\dagger_{zx},d^\dagger_{xy}]$ and $\Delta_{tri}$ is the energy scale for trigonal distortion. The eigenstates are $(\omega=e^{\imath 2\pi/3})$
\begin{eqnarray}
\nonumber
\vert a_{1g}\rangle &= \frac{1}{\sqrt{3}}\left[n_1\vert d_{yz}\rangle + n_2\vert d_{zx}\rangle + n_3\vert d_{xy}\rangle\right],\\
\nonumber
\vert e'_{1g}\rangle &= \frac{1}{\sqrt{3}}\left[\omega n_1\vert d_{yz}\rangle + \omega^2 n_2\vert d_{zx}\rangle + n_3\vert d_{xy}\rangle\right],\\
\vert e'_{2g}\rangle &= \frac{1}{\sqrt{3}}\left[\omega^2 n_1\vert d_{yz}\rangle + \omega n_2\vert d_{zx}\rangle + n_3\vert d_{xy}\rangle\right].
\label{eq_wavefn}
\end{eqnarray}
The trigonal distortion splits the $t_{2g}$ sector into the doubly degenerate $e'_g$ and the non-degenerate $a_{1g}$ with energies $\Delta_{tri}/3$ and $-2\Delta_{tri}/3$ respectively. 

A description based on Hubbard model for the $e'_{g}$ orbitals may be systematically derived starting from the $t_{2g}$ orbitals. This is done in \ref{appen_micro_hamiltonian}. This has the following general form 

\begin{eqnarray}
\nonumber
H'=H^{e'_{g}}_{\rm SO}-\sum_{ ij}\sum_{ M,M'}\sum_\sigma\tilde{t}_{ iM;jM'}e^\dagger_{\rm iM\sigma}e_{\rm iM'\sigma}+\frac{U}{2}\sum_i\sum_{M,M'}\sum_{\sigma\sigma'}e^\dagger_{iM\sigma}e^\dagger_{iM'\sigma'}e_{iM'\sigma'}e_{iM\sigma},\\
\label{eq_hubbard}
\end{eqnarray}
where $e^\dagger_{iM\sigma}$ is the electron creation operator in the $e'_g$ orbital ($M=1,2$) with spin $\sigma(=\uparrow,\downarrow)$; $\tilde{t}_{iM;jM'}$ are the effective hopping amplitudes within the subspace and $U$ is the effective onsite coulomb's repulsion. We note that the Hund's coupling (which arises from the orbital dependence of the Coulomb repulsion) for the $t_{2g}$ orbitals only renormalizes $U$ in this restricted subspace. (See \ref{appen_micro_hamiltonian} for details). 
\subsection{The Projected SO coupling}
 
\begin{figure}
\centering
\includegraphics[scale=0.35]{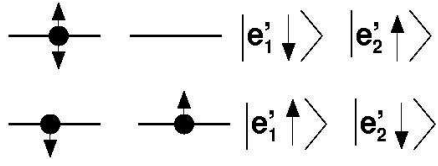}
\caption{The $e'_g$ states split by the SO coupling.}
\label{fig_levels}
\end{figure}

The SO coupling, when projected in the $e'_{g}$ subspace, yields a block diagonal form (see \ref{appen_so_coupling} for details):
\begin{eqnarray}
H^{e'_{g}}_{\rm SO}=-\lambda \hat n\cdot \vec s_i\tau^z_i,
\end{eqnarray}  
where $\vec s_i$ is the spin operator at the site $i$, $\lambda\approx 500\ meV$ is the SO coupling parameter and $\tau^z=+1(-1)$ refers to the $e'_{1g}(e'_{2g})$ orbital. 

Thus the projected SO interaction acts as a ``Zeeman coupling" where the direction of the ``magnetic field" is along the trigonal axis or opposite to it\cite{2009_jin}. Thus it is natural to choose the direction of spin quantization along the axis of trigonal distortion.  This then gives the active atomic orbitals after incorporating the SO coupling. These active orbitals are the Krammer's doublet $\vert e'_{1g},\downarrow\rangle$ and $\vert e'_{2g},\uparrow\rangle$ as shown in Fig. \ref{fig_levels}. 
\subsection{The Spin Hamiltonian}

Hence the low energy physics may be described by considering only the above atomic orbitals. The starting point for the calculations is projection of the Hubbard model (Eq. \ref{eq_hubbard}) in the space spanned by the Krammers's doublet $\vert e'_{1g},\downarrow\rangle$ and $\vert e'_{2g},\uparrow\rangle$. The bandwidth of this projected model is narrow and the effect of the Hubbard repulsion is important. Indeed it can easily render the system insulating. To capture the magnetic order in this Mott insulator, we do a strong-coupling expansion in $\tilde{t}/U$ to get an effective ``pseudo-spin" model in terms of the pseudo-spin-1/2 operators,
\begin{eqnarray}
S^\alpha=\frac{1}{2} e^\dagger_a\rho^\alpha_{ab}e_b,
\end{eqnarray}
where, $\rho^\alpha\ (\alpha=x,y,z)$ are the Pauli matrices and $a,b=(e'_{g1};\downarrow),(e'_{g2};\uparrow)$. The ``pseudo-spin" Hamiltonian has the following form up to the quadratic order:

\begin{eqnarray}
\nonumber
&H=\sum_{\langle ij\rangle} J^{(1)}_{ij} \vec S_i\cdot\vec S_j +\sum_{\langle\langle ij\rangle\rangle}J^{(2)}_{ij}\vec S_i\cdot\vec S_j +\sum_{\langle\langle\langle ij\rangle\rangle\rangle}J^{(3)}_{ij}\vec S_i\cdot\vec S_j\\
&+\sum_{\langle ij\rangle}J^{(z1)}_{ij}S^z_iS^z_j+\sum_{\langle\langle ij\rangle \rangle} J^{(z2)}_{ij}S^z_iS^z_j +\sum_{\langle\langle\langle ij\rangle \rangle\rangle} J^{(z3)}_{ij}S^z_iS^z_j.
\label{eq_hamiltonian}
\end{eqnarray}
Here $\langle ij\rangle$, $\langle\langle ij\rangle\rangle$ and $\langle\langle\langle ij\rangle\rangle\rangle$ refer to summation over first, second and third nearest neighbours (NNs) respectively.

The different exchange couplings are given in terms of the underlying parameters of the Hubbard model as

\begin{eqnarray}
J_{ij}^{(z\alpha)}&=&\frac{8}{U}(T^{(z\alpha)}_{ij})^2, \nonumber \\
J_{ij}^{(\alpha)} &=&\frac{4}{U}\left[(T^{(0\alpha)}_{ij})^2-(T^{(z\alpha)}_{ij})^2\right]
=J_{ij}^{(0\alpha)}-\frac{1}{2}J_{ij}^{(z\alpha)}
\label{eq_exchange_couplings}
\end{eqnarray}
where $\alpha=1,2,3$ denotes that $ij$ are first, second or third NNs, respectively and the last expression defines $J_{ij}^{(0 \alpha)}$. $T^{(0\alpha)}_{ij}$ and $T^{(z\alpha)}_{ij}$ are given in terms of the hopping amplitudes (e.g. $t^{xy;yz}_{ij}$ from the overlap of $xy$ and $yz$ orbitals) of the $t_{2g}$ orbitals as (details are given in \ref{appen_parameters}).
\begin{eqnarray}
\nonumber
T^{(0\alpha)}_{ij}=&&\frac{1}{3}\left[\left(t^{yz;yz}_{ij}+ t^{xz;xz}_{ij}+t^{xy;xy}_{ij}\right)\right]\\
\nonumber
&&-\frac{1}{6}\left[\left(n_1\left(t^{xz;xy}_{ij}+t^{xy;xz}_{ij}\right)+n_2\left(t^{xy;yz}_{ij}+t^{yz;xy}_{ij}\right)+n_3\left(t^{yz;xz}_{ij}+t^{xz;yz}_{ij}\right)\right)\right]\\
\end{eqnarray}
\begin{eqnarray}
\nonumber
T^{(z\alpha)}_{ij}=\frac{1}{2\sqrt{3}}\left[n_1\left(t^{xz;xy}_{ij}-t^{xy;xz}_{ij}\right)+n_2\left(t^{xy;yz}_{ij}-t^{yz;xy}_{ij}\right)+n_3\left(t^{yz;xz}_{ij}-t^{xz;yz}_{ij}\right)\right]\\
\end{eqnarray}

Before moving on to the details of the spin Hamiltonian, we note that, on projecting to the subspace of $\vert e'_{1g},\downarrow\rangle$ and $\vert e'_{2g},\uparrow\rangle$, the spin and orbitals are no longer independent. Instead at every site there is a pseudo-spin-$1/2$ degree of freedom where the spin is {\em locked} to the orbital wave function. This, we refer to as {\em spin-orbital locking}.

\section{Application to N${\rm\bf a}_2{\rm\bf Ir}$O$_3$}
\label{sec_meanfield}
We now apply the above results to the case of Na$_2$IrO$_3$. The early X-Ray diffraction experiments\cite{2010_singh} suggested a a monoclinic $C2/C$ structure for the compound and distorted IrO$_6$ octahedra. However, more recent experiments see a better match for the X-Ray diffraction data with the space group $C2/m$. \cite{2012_choi,2012_ye} They also unambiguously confirm the presence of uniform trigonal distortion of the IrO$_6$ octahedra. However, the magnitude of such distortion is not clear at present. Further, experimental measurements suggest: (1) The magnetic transition occurs at $T_N=15K$ while the Curie-Weiss temperature is about $\Theta_{CW} \approx -116K$. This indicates presence of frustration. (2) The high temperature magnetic susceptibility is anisotropic; the in-plane and out-of-plane susceptibilities are different. This may be due to a trigonal distortion of the IrO$_6$ octahedra \cite{2010_singh}. (3) The magnetic specific heat is suppressed at low temperatures.\cite{2010_singh} (4) Recent resonant X-ray scattering experiment \cite{2011_liu} suggests that the magnetic order is collinear and have a 4-site unit cell. (5) The magnetic moments have a large projection on the $a$-axis of the monoclinic crystal \cite{2011_liu}. (6) A combination of these experimental findings and density functional theory (DFT) calculations strongly suggest that a `zig-zag' pattern for the magnetic moments, as shown in Fig. \ref{fig_zig_zag} in the ground state {\cite{2011_liu}}, which has since been verified independently by two groups using Neutron scattering \cite{2012_choi,2012_ye}.

 Taking these phenomenological suggestions, we try to apply the above calculations to the case of Na$_2$IrO$_3$. At the outset, we must note that, in the above derivation of the spin Hamiltonian we have assumed that the trigonal distortion to be the largest energy scale followed by the SO coupling. While this extreme limit of projecting out the $a_{1g}$ orbitals most likely is not true for Na$_2$IrO$_3$. However, we expect the real ground state to be adiabatically connected to this limit. With this in mind, we now consider the case of Na$_2$IrO$_3$.

\begin{figure}
\centering
\includegraphics[scale=0.3]{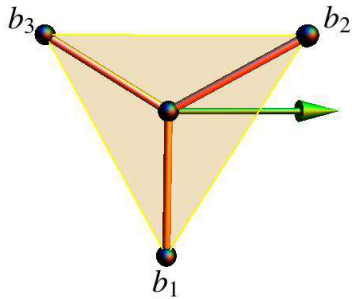}
\caption{Section of a honeycomb lattice (shaded in yellow). Ir sites (black) are connected by bonds (orange). The green arrow is the the $[-1,-1,1]$ direction of trigonal distortion that makes an angle of about $19^\circ$ with the plane of the lattice pointing inside the plane.}
\label{fig_exchanges}
\end{figure}

Clearly, the exchanges (Eq.~\ref{eq_exchange_couplings}) depend both on the direction of the bond and the direction of the trigonal distortion. So it is important to ask about the direction of the latter. Comparing the crystallographic axes of Na$_2$IrO$_3$, we find that the direction $[1,1,1]$ is perpendicular to the honeycomb plane while the other three directions make an acute angle to it.  In the monoclinic $C2/m$ structure, uniform trigonal distortion in these four directions may not cost the same energy. In experiments {\cite{2011_liu}}, the moments are seen to point along the $a$-axis of the monoclinic crystal which is parallel to the honeycomb plane. This, along with the fact that the magnetic moment in our model is in the direction of $\hat n$ (explained below) seems to suggest that $\hat n=\frac{1}{\sqrt{3}}[-1,-1,1]$ is chosen in the compound (see Fig.~\ref{fig_exchanges}). In the absence of a better theoretical understanding of the direction of the trigonal distortion, we take this as an input from the experiments.  

To identity different hopping paths (both direct and indirect), we consider various overlaps (see \ref{appen_parameters}) and find, while $J^{(3z)}=0$, $J^{(1z)}\neq J^{(2z)}\neq 0$ are approximately (spatially-)isotropic and antiferromagnetic. For the exchanges of the Heisenberg terms, both $J^{(2)}$ and $J^{(3)}$ are antiferromagnetic and isotropic (both of them result from indirect hopping mediated by the Na s-orbitals and are expected to be comparable). For the NN Heisenberg exchanges, the couplings are antiferromagnetic, but, much more spatially anisotropic. We find that for the chosen direction of the trigonal distortion, the coupling along one of the NN exchanges ($J^{(1)}$) ({\em viz} $b_1$ in Fig. \ref{fig_exchanges}) is different from the other two neighbours ($\tilde{J}^{(1)}$)($b_2$ and $b_3$ in Fig.~\ref{fig_exchanges}).

\begin{figure}
\centering
\includegraphics[scale=0.5]{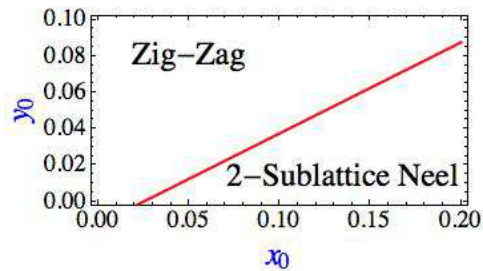}
\caption{Mean field phase diagram for Eq. \ref{eq_hamiltonian}. The two axes are: $x_0 = \frac{\tilde J^{(1)}}{J^{(1)}}$; $y_0 = \frac{J^{(2)}}{J^{(1)}}$, where $J^{(1)}(\tilde{J}^{(1)})$ are related to the strong(weak) NN exchange and $J^{(2)}$ is the $2^{nd}$ and $3^{rd}$ neighbour exchange (see Eq. \ref{eq_exchange_couplings}). We take the Ising anisotropy to be $5\%$ of $J^{(1)}$. Note that, due to Ising anisotropies, one has zig-zag order at $y_0=0$.} 
\label{fig_pd}
\end{figure}

\subsection{Mean-Field Theory and Magnetic Order}

We now consider the mean field phase diagram for the above anisotropic spin Hamiltonian. For $J^{(1)}$ being the largest energy scale, the classical ground state for the model can be calculated within mean-field theory as a function of $x_0=\tilde{J}^{(1)}/J^{(1)}$ and $y_0=J^{(2)}/J^{(1)}$ (we have taken $J^{(2)}=J^{(3)}$).  A representative mean-field phase diagram is shown in Fig.~\ref{fig_pd}. It shows a region of the parameter-space where the zig-zag order is stabilized {\cite{2009_chkim}}. The effect of the Ising anisotropies $J^{(1z)}$ and $J^{(2z)}$ is to pin the magnetic ordering along the $z$-direction of the pseudo-spin quantization which is also the direction of the trigonal distortion $\hat n$. They also gap out any Goldstone mode that arises from the ordering of the pseudo-spins. The latter results in the exponential suppression of the specific heat at low temperatures. The other competing phase with a collinear order is the regular two-sublattice Neel phase. 

The nature of the ground states may be understood from the following arguments. In the presence of the $\hat n$ in $[-1,-1,1]$ direction, the NN exchange coupling becomes anisotropic. When it is strong in one direction ($J^{(1)}$) and weak in two other directions ($\tilde{J}^{(1)}$), for the bonds where the NN coupling becomes weak, the effects of the small second and third neighbour interactions become significant. Since the latter interactions are antiferromagnetic, they prefer anti-parallel alignment of the spins. As there are more second and third neighbours, their cumulative effect can be much stronger. This naturally leads to the zig-zag state. The NN antiferromagnetic interactions on the weaker bonds compete with the antiferromagnetic second and third neighbour interactions and frustrates the magnet. This suppresses the magnetic ordering temperature far below the Curie-Weiss temperature.

\subsection{The spectrum for Spin-orbital waves} 
 
 The low energy excitations about this magnetically ordered zig-zag state are gapped {\em spin-orbital} waves. Signatures of such excitations may be seen in future resonant X-Ray scattering experiments. It is important to note that this ``pseudo-spin" waves actually contain both orbital and the spin components due to the spin-orbital locking.
 
\begin{figure}
\centering
\subfigure[]{
\includegraphics[scale=0.4]{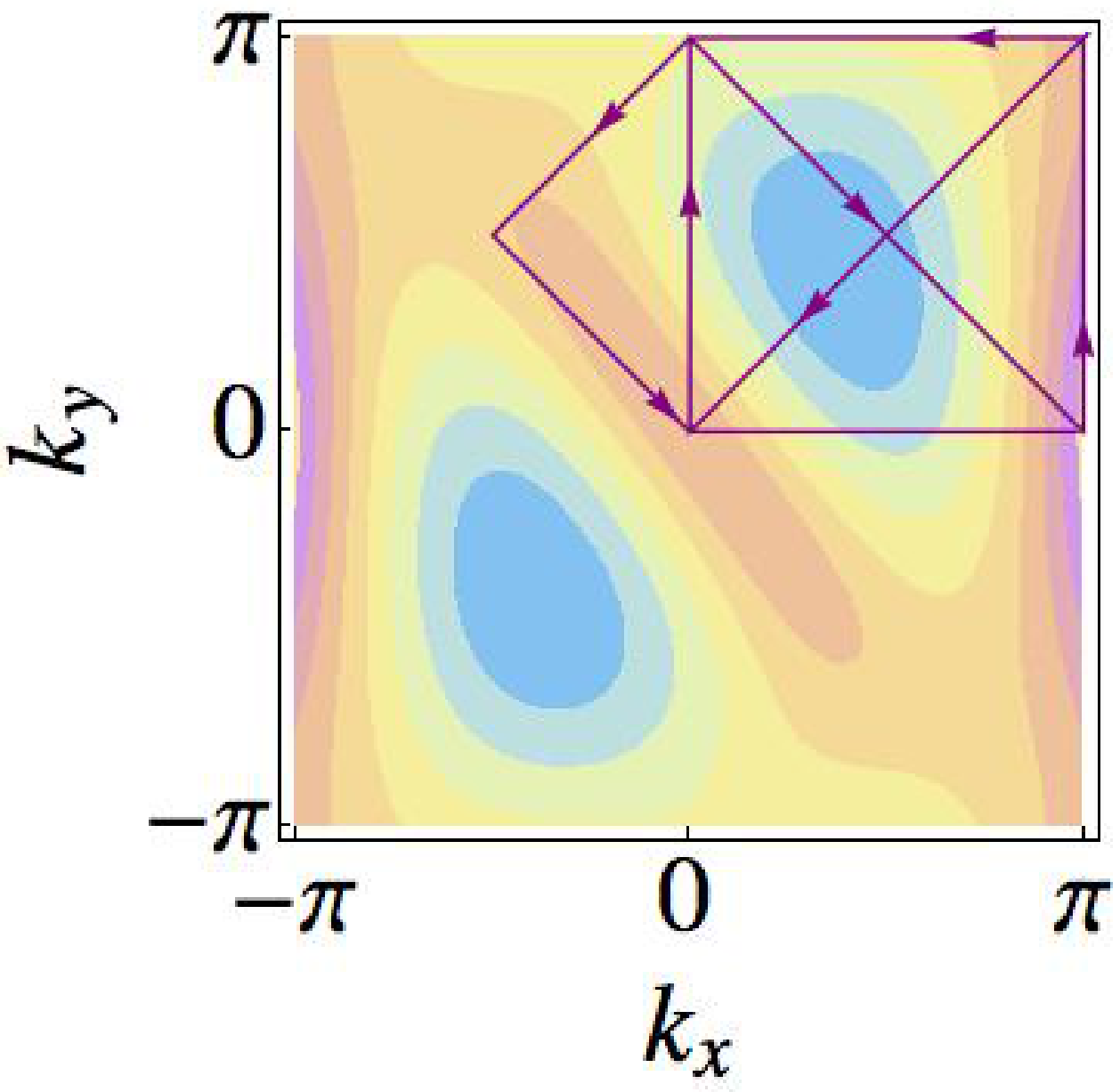}
\includegraphics[scale=0.4]{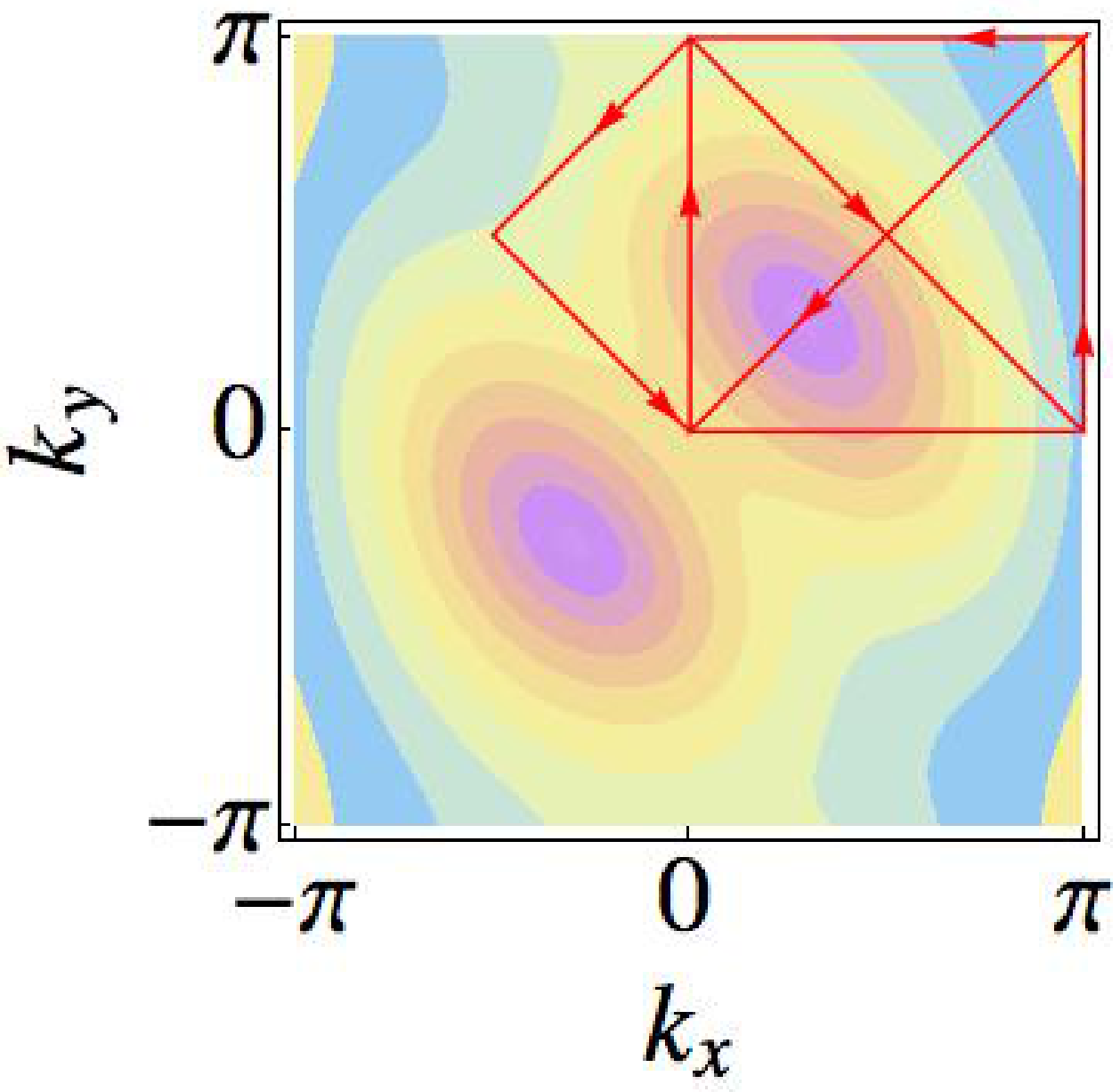}
\label{fig_sw1}
}
\subfigure[]{
\includegraphics[scale=0.3]{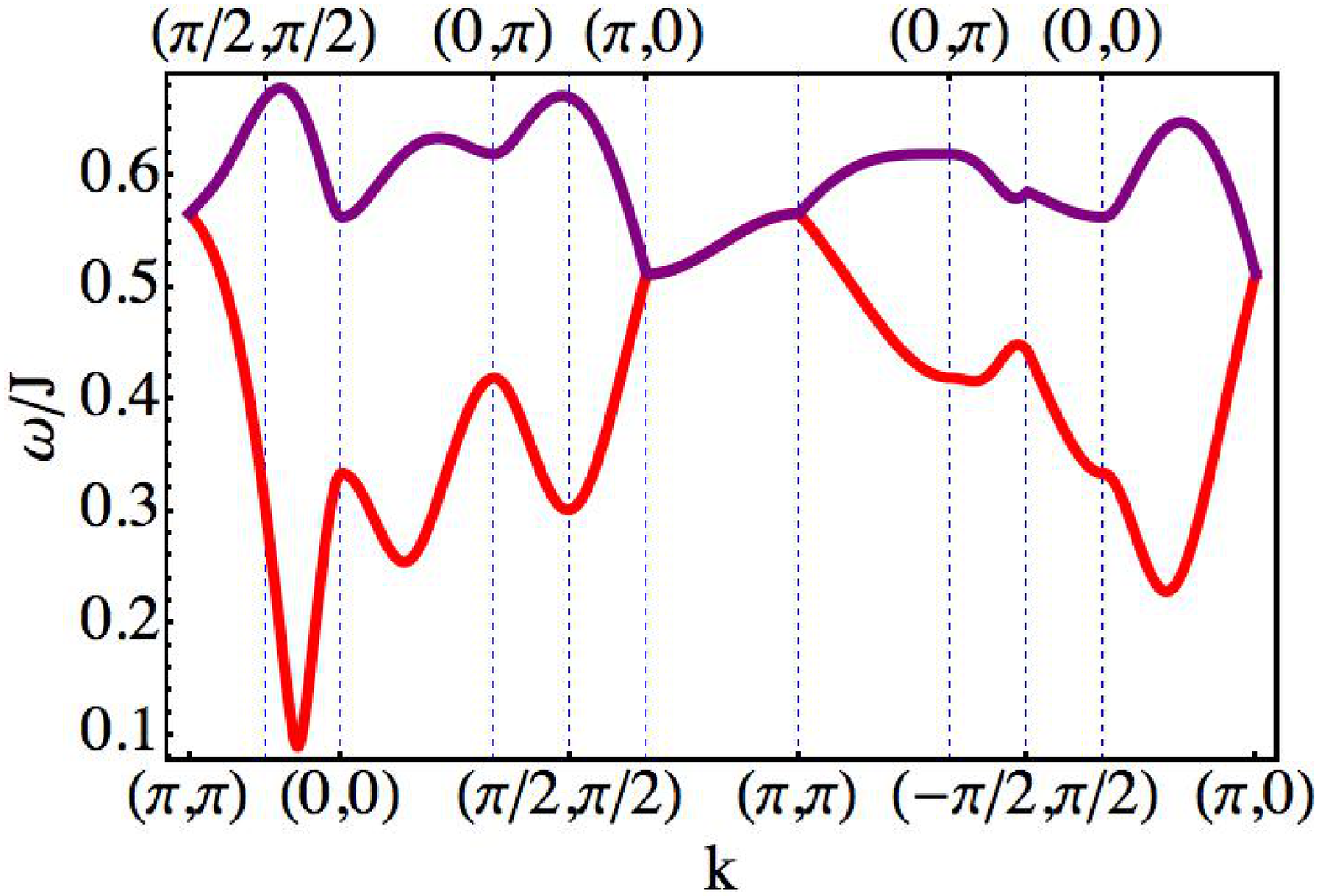}
\label{fig_sw2}
}
\caption{The ``pseudo-spin" wave spectrum (contours of both the bands are shown in (a) and a section is shown in (b)). The values used for the parameters are same as that used for the calculation of the mean field phase diagram (Fig. \ref{fig_pd}). We note that, as expected, the spectrum is gapped.}
\end{figure}

We calculate the dispersion of such spin-orbital waves to quadratic order using the well-known Holstein-Primakoff methods. The details are discussed in \ref{appen_HP_boson}. A representative spin wave spectrum in the zig-zag phase is shown in Fig.~\ref{fig_sw1} and \ref{fig_sw2}. The spectrum is gapped and the bottom of the spin-wave dispersion has some characteristic momentum dependence.

\subsection{Experimental Implications}

Apart from the already discussed exponential suppression of low temperature magnetic specific heat, the above calculation predicts an interesting feature in the magnetic susceptibility. The relation between the magnetic moment and the pseudo-spins is 
\begin{eqnarray}
\vec M_i=-4\mu_B \hat n S^z_i,
\label{eq_mag}
\end{eqnarray}
where $\mu_B$ is the Bohr magneton. This follows from the twin facts that, in $e'_g$ subspace, the angular momentum transverse to $\hat n$ is quenched and the spins are locked to the orbitals with the axis of quantization being $\hat n$ in our pseudo-spin sector (see \ref{appen_zeeman}). Thus, the magnetization is sensitive to the $z$-component of the pseudo-spin (the direction of which is shown in Fig. \ref{fig_zig_zag}). Indeed the magnetization has the largest projection along the $a$-axis of the monoclinic crystal. This was seen in experiments \cite{2011_liu} and was the motivation for choosing the $[-1,-1,1]$ direction for the trigonal distortion. Along two other axes $[-1,1,-1]$ and $[1,-1,-1]$, a large component of in-plane magnetization exists, but in different directions. Finally the direction $[1,1,1]$ is perpendicular to the honeycomb plane and leads to magnetization in the same [1,1,1] direction. While this does not appear to be the case for Na$_2$IrO$_3$, this may be more relevant for the less-distorted compound Li$_2$IrO$_3$ (see below). Eq.~\ref{eq_mag} suggests that the magnetic susceptibility is highly anisotropic and depends on the cosine of the angle between the direction of magnetic field and $\hat n$. Indeed signatures of such anisotropy have been already seen in experiments \cite{2010_singh}. We emphasize that within this picture, the in-plane susceptibility also varies with the direction of the magnetic field. So the ratio $\chi_\perp/\chi_{\Vert}$ can be lesser or greater than $1$. The current experiments\cite{2010_singh} does not tell the in-plane direction of magnetic field and hence we cannot comment on the ratio presently.  However, the above picture is strictly  based on atomic orbitals. One generally expects that there is also hybridization of the Ir $d$-orbitals with the oxygen $p$-orbitals. Such hybridization will contribute to a non-zero isotropic component to the susceptibility \cite{2010_chen}. Also, as remarked earlier, in the actual compound, the SO coupling scaling may not be very small compared to the trigonal distortion limit scale. Additional perturbation coming from the mixing with the $a_{1g}$ orbitals will also contribute to decrease the anisotropy of the susceptibility.

\section{Summary and Conclusion}
\label{sec_summary}
In this paper, we have studied the effect of trigonal distortion and SO coupling and applied it to the case of the honeycomb lattice compound Na$_2$IrO$_3$. We find that, in the limit of large trigonal distortion and SO coupling, a pseudo-spin-$1/2$ degree of freedom emerges. Low energy Hamiltonian, in terms of this pseudo-spin gives a `zig-zag' magnetic order as seen in the recent experiments on Na$_2$IrO$_3$. We have also calculated the low energy spin-wave spectrum and elucidated various properties of the compound that has been observed in experiments. The pseudo-spin couples the physical spin and the orbitals in a non-trivial manner, signatures of this may be seen in future inelastic X-ray resonance experiments probing the low energy excitations.

While very recent experiments \cite{2012_choi,2012_ye} clearly indicate presence of trigonal distortions, their magnitude is yet not confirmed. On the other hand, the only available numerical estimate of the energy scale for trigonal distortion comes from the DFT calculations by Jin {\em et al.} \cite{2009_jin} (based on $C2/C$ structure). It suggests $\Delta_{tri}\approx 600\ meV$. While, it is not clear if such a large value is in confomity with the experiments, at present, the detection of trigonal distortion in experiments is highly encouraging from the perspective of the present calculations.  

In these lights of the above calculations, it is tempting to predict the case of Li$_2$IrO$_3$ where recent experiments suggest a more isotropic honeycomb lattice \cite{2011_singh,2011_takagi}.  A possibility is that sizeable trigonal distortion is also present in Li$_2$IrO$_3$ (so that the above discussion holds), but, the axis is perpendicular to the plane. What may be the fallouts in such a case ? Our present analysis would then suggest that the antiferromagnetic exchanges are isotropic and equally strong for the three NNs. This would develop 2-sublattice Neel order in the pseudo-spins with the magnetic moments being perpendicular to the plane. Also the further neighbour exchanges are rather weak (compared to Na$_2$IrO$_3$) and hence frustration is quite small. Indeed recent experiments see ordering very close to the Curie-Weiss temperature, the later being calculated from the high temperature magnetic susceptibility data \cite{2011_takagi,2011_singh}. However, present experiments do not rule out the possibility of small or no trigonal distortions in Li$_2$IrO$_3$, in which case the limit of HK model \cite{2010_chaloupka,2011_reuter} may be
appropriate.

\acknowledgements

We acknowledge useful discussion with H. Gretarsson, R. Comin, S. Furukawa, H. Jin, C. H. Kim, Y.-J. Kim, W. Witczak-Krempa, H. Takagi. YBK thanks the Aspen Center for Physics, where parts of the research were done. This work was supported by the NSERC, Canadian Institute for Advanced Research, and Canada Research Chair program.

\appendix
\section{The microscopic model for Na$_2$IrO$_3$}
\label{appen_micro_hamiltonian}
The generic Hubbard model (for the $t_{2g}$ orbitals) including the trigonal distortions, Hund's coupling and the SO coupling is
\begin{eqnarray}
\nonumber
H&=-\lambda\sum_{i}\vec l_i\cdot \vec s_i + H^{t_{2g}}_{tri} + \sum_{ij}\sum_{mm'}\sum_{\sigma\sigma'}\left(t_{ij}^{m;m'}d^\dagger_{im\sigma}d_{jm'\sigma'}\right)\\
&+\frac{1}{2}\sum_i\sum_{mm'}\sum_{\sigma\sigma'}U_{mm'}d^\dagger_{im\sigma}d^\dagger_{im'\sigma'}d_{im'\sigma'}d_{im\sigma}.
\end{eqnarray}
Here $m,m'=yz,xz,xy$ and $\sigma=\uparrow,\downarrow$ and $H^{t_{2g}}_{tri}$ is given by Eq. \ref{eq_tri_ham}. We note that the hopping is diagonal in spin space and in the cubic harmonic basis all hopping are real. Also, the hopping contain both the direct and indirect (through Oxygen and Sodium) paths. We have taken Hund's coupling into account through $U_{mm'}$, though this is expected to be small in $5d$ transition metals. To a very good approximation the form of $U_{mm'}$ is given by
\begin{eqnarray}
U_{mm'}\equiv\left[\begin{array}{ccc}
U_0 & U_0-J_H & U_0-J_H\\
U_0-J_H & U_0 & U_0-J_H\\
U_0-J_H & U_0-J_H & U_0\\
\end{array}\right],
\end{eqnarray}
where the basis is given, as before, by $\Psi_i^\dagger=[d^\dagger_{yz},d^\dagger_{zx},d^\dagger_{xy}]$. $U_0$ and $J_H$ are the intra orbital Coulomb repulsion and Hund's coupling term respectively.

The transformation between the operators in the trigonal basis, $\Phi^\dagger=\left[a^\dagger_{1g},e'^\dagger_{1g},e'^\dagger_{2g}\right]$,  and  $t_{2g}$ basis, $\Psi^\dagger=\left[d^\dagger_{yz},d^\dagger_{zx},d^\dagger_{xy}\right]$, is given by $\Psi_{m}=T_{m,M}\Phi_{M}$. The transformation matrix is given by
\begin{eqnarray}
T_{m,M}=\frac{1}{\sqrt{3}}\left[\begin{array}{ccc}
n_1 & n_1\omega  & n_1\omega^2 \\
n_2 & n_2\omega^2 & n_2\omega \\
n_3 & n_3 & n_3 \\
\end{array}\right].
\end{eqnarray}
The transformations for the hopping amplitudes and repulsion term are then given by 
\begin{eqnarray}
\nonumber
\tilde t_{iM;jM'}&=\sum_{m,m'}T^{*}_{m,M}t_{ij}^{m;m'}T_{m',M'};\\\tilde{U}_{M_1M_2}&=\sum_{m,m'}U_{mm'}\left(T^*_{mM_1}T_{mM_1}\right)\left(T_{m'M_2}^*T_{m'M_2}\right).
\end{eqnarray}
Notice that there are contributions to $t_{ij}^{m;m'}$ from both direct and indirect exchanges for the first, second and third neighbours, as confirmed from the DFT calculations by H. Jin {\em et al.} {\cite{2009_jin}}. These show that there are contributions from both direct and indirect hoppings for the first, second and third nearest neighbours. Projecting them into the $e'_g$ orbitals we get the effective hopping amplitudes which are then used in Eq. \ref{eq_hubbard}. As for the Coulomb repulsion term, we find that it has the following form
\begin{eqnarray}
\tilde{U}_{M_1M_2}=U\left[\begin{array}{ccc}
1 & 1 & 1\\
1 & 1 & 1\\
1 & 1 & 1\\
\end{array}\right],
\end{eqnarray} 
where $U=U_0-2J_H/3$. This form is then used in Eq. \ref{eq_hubbard}. The reason for this special form of $\tilde{U}_{M_1M_2}$  lies in the fact that the $e'_g$ orbitals have equal weight of the three $t_{2g}$ orbitals (see the wave functions in Eq. \ref{eq_wavefn}). 
\section{Projection of Spin-Orbit coupling to the $e'_g$ subspace}
\label{appen_so_coupling}
The SO coupling, when projected to the $t_{2g}$ orbitals give
\begin{eqnarray}
H^{t_{2g}}_{\rm SO}=-\lambda \vec l\cdot \vec s,
\end{eqnarray}
where $\vec l$ is a $l=1$ angular momentum operator. We can re-write the $t_{2g}$ cubic harmonics in terms of the spherical harmonics of the effective $l=1$ angular momentum operator. These are given by:
\begin{eqnarray}
\nonumber
\vert d_{yz}\rangle &= \frac{1}{\sqrt{2}}\left[\vert 1,-1\rangle-\vert 1,+1\rangle\right];\\
\nonumber
\vert d_{zx}\rangle &= \frac{\imath}{\sqrt{2}}\left[\vert 1,-1\rangle+\vert 1,-1\rangle\right];\\ 
\vert d_{xy}\rangle &= \vert 1,0\rangle
\end{eqnarray}
The projector for the $e'_g$ space is: $P^{e'_g}=\vert e'_{1}\rangle \langle e'_{1}\vert + \vert e'_{2}\rangle \langle e'_{2}\vert$. It turns out that $\vec l\cdot\vec s$ is block diagonal in this subspace. Hence
\begin{eqnarray}
\vec l\cdot \vec s=\vert e'_{1}\rangle \langle e'_{1}\vert\vec l\cdot\vec s\vert e'_{1}\rangle \langle e'_{1}\vert + \vert e'_{2}\rangle \langle e'_{2}\vert\vec l\cdot\vec s\vert e'_{2}\rangle \langle e'_{2}\vert
\end{eqnarray}

Making the ``gauge" choice we get
\begin{eqnarray}
\langle e'_{1}\vert\vec l\cdot\vec s\vert e'_{1}\rangle = \hat n\cdot \vec s;\ \ \ \ \langle e'_{2}\vert\vec l\cdot\vec s\vert e'_{2}\rangle =-\hat n\cdot \vec s
\end{eqnarray} 
\section{The hopping parameters}
\label{appen_parameters}

\begin{figure}
\centering
\subfigure[]{
\includegraphics[scale=0.3]{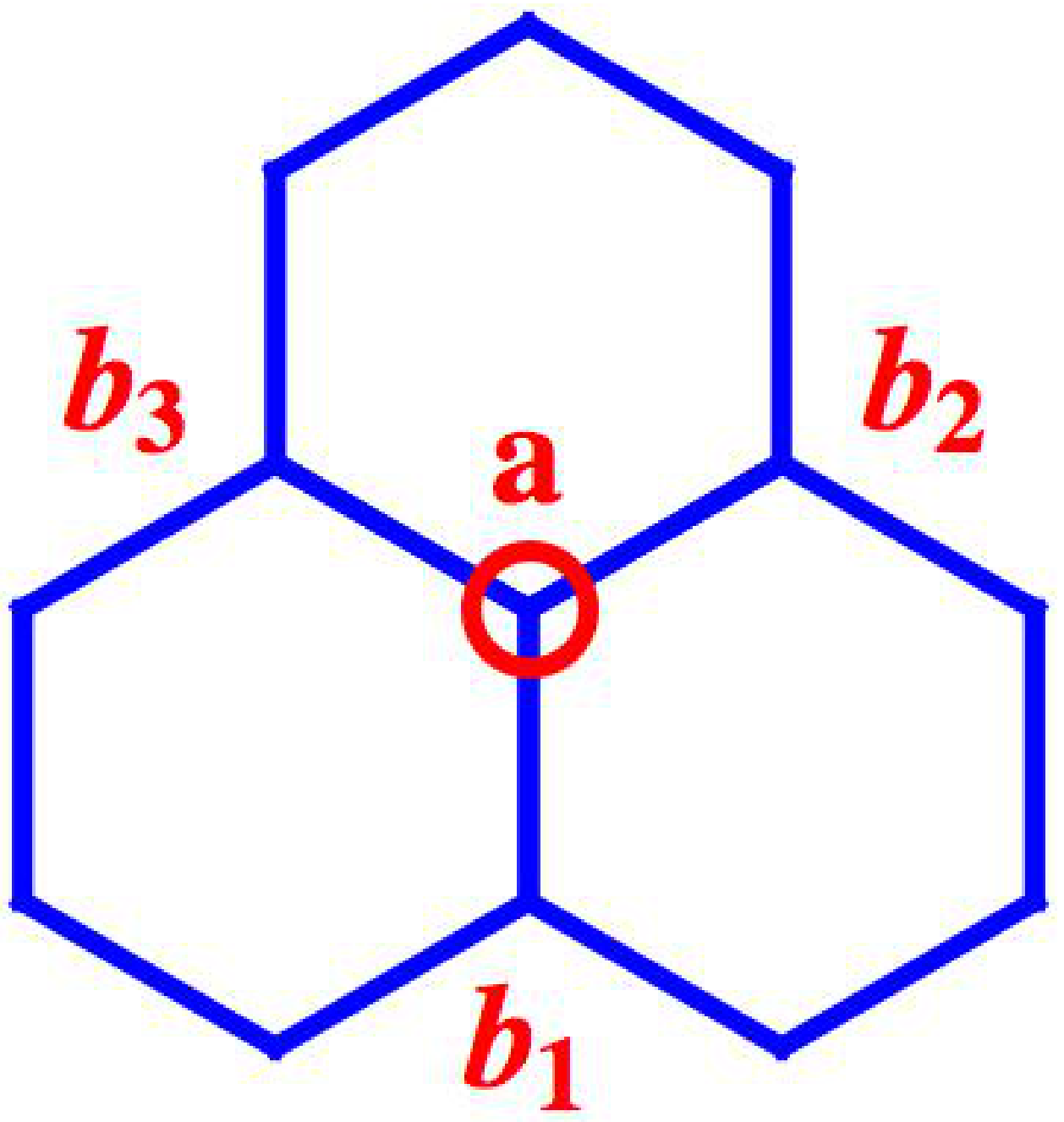}
\label{lattice}
}
\subfigure[]{
\includegraphics[scale=0.4]{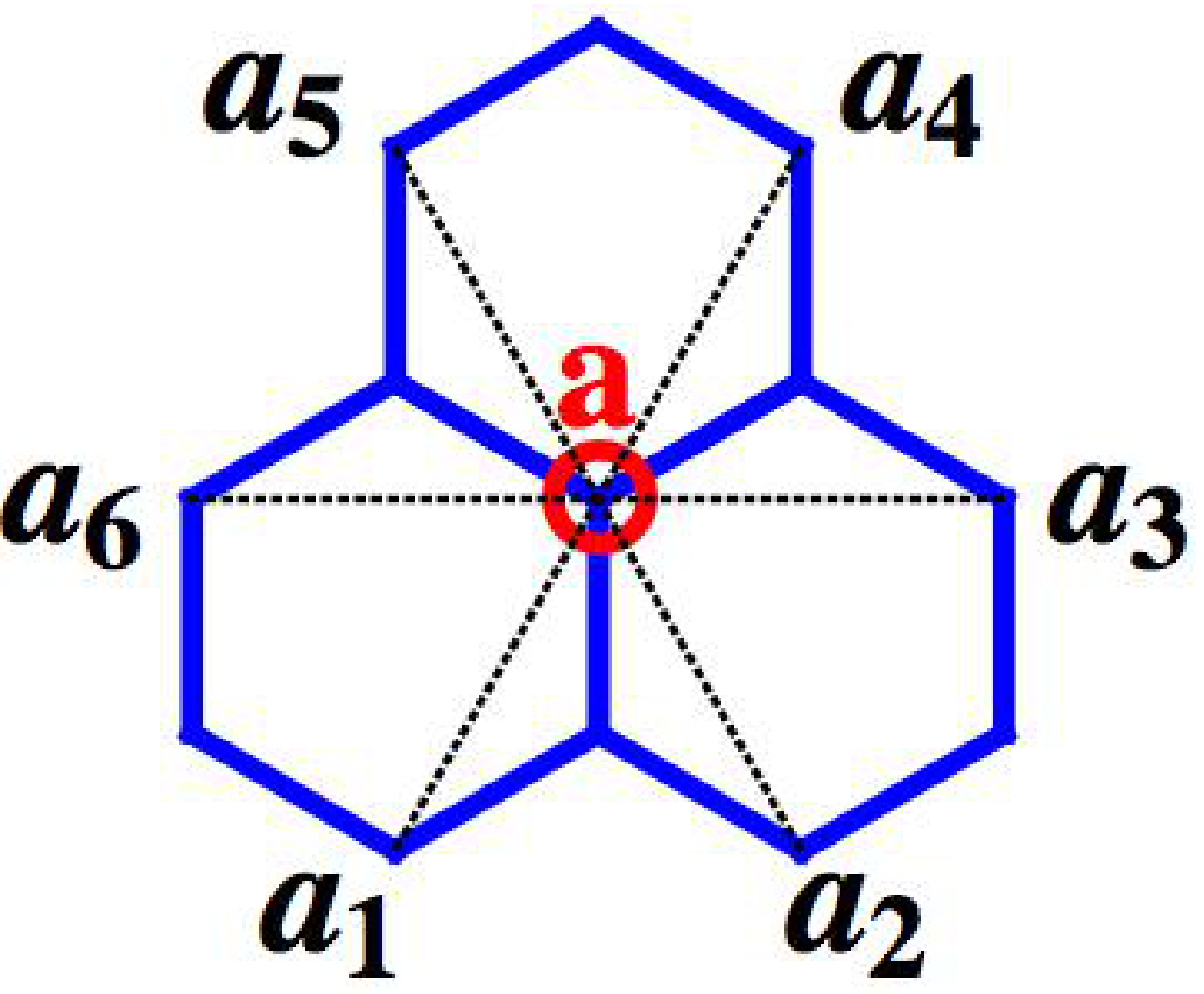}
\label{n3lattice}
}
\subfigure[]{
\includegraphics[scale=0.4]{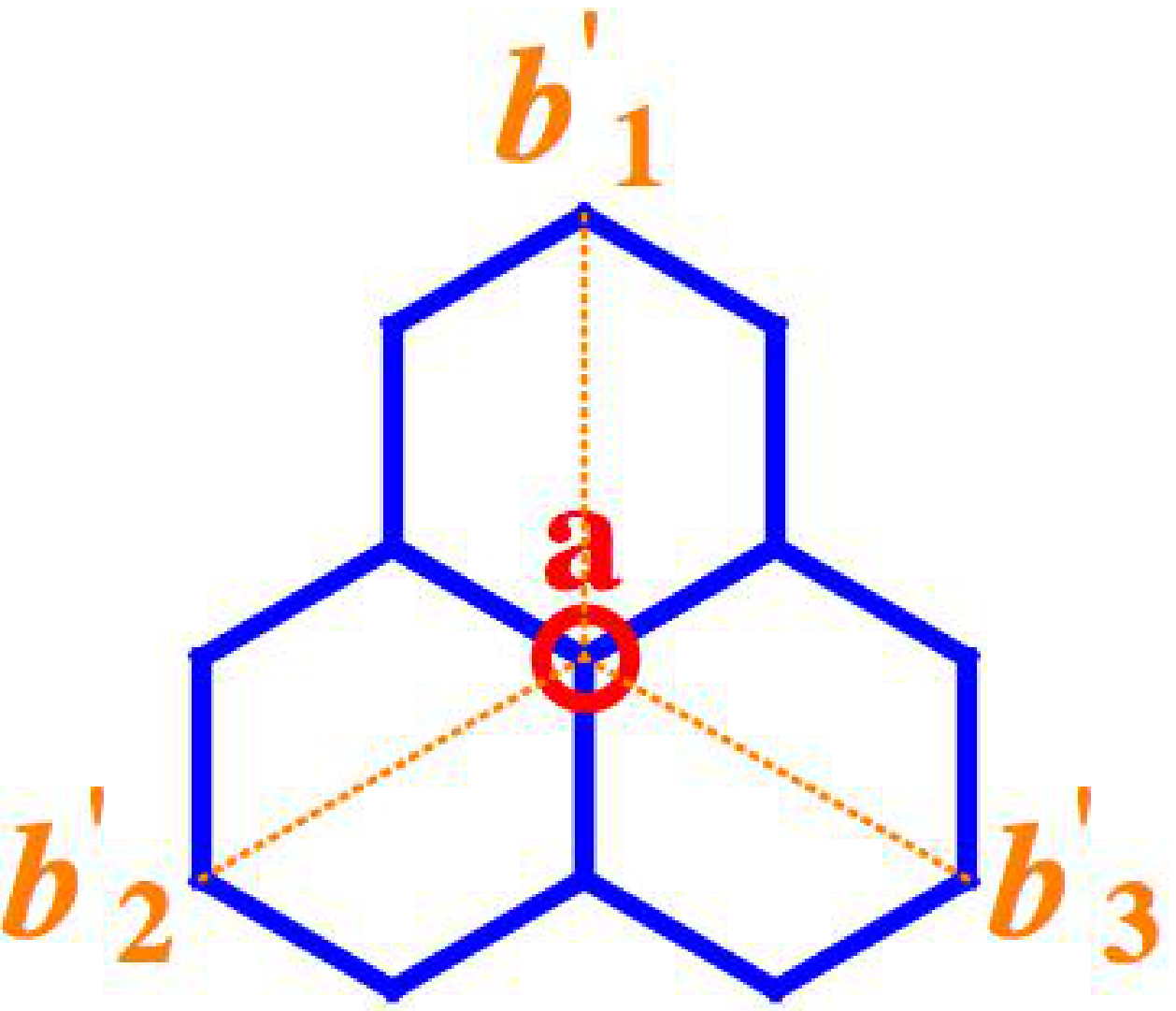}
\label{n4lattice}
}
\caption{The 3 nearest neighbours (a),  six 2$^{nd}$ nearest neighbours (b) and three 3$^{rd}$ nearest neighbours (c) of the central site. The nomenclature has been used to label the hoppings.}
\end{figure}

\subsection{Nearest neighbours}
The nearest neighbours are shown in Fig. \ref{lattice}. There are two different processes contributing to the hopping.: 1) the direct hopping between the Ir atoms and 2) the indirect hopping between the Ir atoms mediated by the oxygen atoms. In presence of the trigonal distortion which has a component along the honeycomb plane (like in this case $[-1,-1,1]$) the magnitudes of the different hopping parameters are different in different directions (for both direct and indirect hopping). The results are shown in Table \ref{tab_nn}. 

\begin{table}
\centering
\subtable[NN:$t_{am;b_1m'}$]{
\begin{tabular}{|c|c|c|c|}\hline
$ _{m'}\backslash ^m$ & $d_{xy}$ & $d_{yz}$ & $d_{zx}$ \\ \hline
$d_{xy}$ & $t_{dd1}(b_1)$ & - & - \\ \hline
$d_{yz}$ & - & $t_{dd2}(b_1)$ & $-t_{dd2}(b_1) + t_0(b_1)+\Delta_1(b_1)$ \\ \hline
$d_{zx}$ & - & $-t_{dd2}(b_1) + t_0(b_1)-\Delta_1(b_1)$ & $t_{dd2}(b_1)$\\ \hline
\end{tabular}
}
\subtable[NN:$t_{am;b_2m'}$]{
\begin{tabular}{|c|c|c|c|}\hline
$ _{m'}\backslash ^m$ & $d_{xy}$ & $d_{yz}$ & $d_{zx}$ \\ \hline
$d_{xy}$ & $t_{dd2}(b_2)$ & - & $-t_{dd2}(b_2)+t_0(b_2)+\Delta_1(b_2)$ \\ \hline
$d_{yz}$ & - & $t_{dd1}(b_2)$ & - \\ \hline
$d_{zx}$ & $-t_{dd2}(b_2)+t_0(b_2)-\Delta_1(b_2)$ & - & $t_{dd2}(b_2)$\\ \hline
\end{tabular}
}
\subtable[NN: $t_{am;b_3m'}$]{
\begin{tabular}{|c|c|c|c|}\hline
$ _{m'}\backslash ^m$ & $d_{xy}$ & $d_{yz}$ & $d_{zx}$ \\ \hline
$d_{xy}$ & $t_{dd2}(b_3)$ & $-t_{dd2}(b_3)+t_0(b_3)+\Delta_1(b_3)$ & -\\ \hline
$d_{yz}$ & $-t_{dd2}(b_3)+t_0(b_3)-\Delta_1(b_3)$ & $t_{dd2}(b_3)$ & - \\ \hline
$d_{zx}$ & - & - & $t_{dd1}(b_3)$\\ \hline
\end{tabular}
}
\caption{The hopping paths (both direct and indirect) in the $t_{2g}$ basis.}
\label{tab_nn}
\end{table}

We shall make an approximation here. We shall leave out the directional dependence of the magnitudes on the direction. The argument is that the essential directional dependence due to the trigonal distortion has been taken care of by the parameter $\Delta_1$. When the DFT {\cite{2009_jin}} results are used to find the tight-binding parameters {\cite{2009_chkim}}, it is found that (they use $\Delta_1=0$) (here $t_{dd1}$ and $t_{dd2}$ are direct hopping and $t_0$ is the indirect hopping respectively.) $t_{dd1}=-0.5\ eV; t_{dd2}=0.15\ eV;t_0 = 0.25\ eV.$

Performing the transformation to the $e'_{g}$ basis, we have 
\begin{eqnarray}
\nonumber
T^{(01)}_{ab_1}&=\frac{1}{3}\left[t_{dd1} + 2 t_{dd2} + (t_{dd2}-t_0)n_3\right],\\
\nonumber
T^{(02)}_{ab_2}&=\frac{1}{3}\left[t_{dd1} + 2 t_{dd2} + (t_{dd2}-t_0)n_1\right],\\
T^{(03)}_{ab_3}&=\frac{1}{3}\left[t_{dd1} + 2 t_{dd2} + (t_{dd2}-t_0)n_2\right].
\end{eqnarray}
and
\begin{eqnarray}
\nonumber
T^{(z1)}_{ab_1}&=-\frac{\Delta_1}{\sqrt{3}}n_3\\
\nonumber
T^{(z1)}_{ab_2}&=-\frac{\Delta_1}{\sqrt{3}}n_1\\
 \ T^{(z1)}_{ab_3}&=-\frac{\Delta_1}{\sqrt{3}}n_2
\end{eqnarray}
Hence,
\begin{eqnarray}
\nonumber
J^{(0)}_{ab_1}&=\frac{4}{3U}\left[\frac{1}{3}\left[t_{dd1} + 2 t_{dd2} + (t_{dd2}-t_0)n_3\right]^2-(\Delta_1)^2\right], \\
\nonumber
J^{(0)}_{ab_2}&=\frac{4}{3U}\left[\frac{1}{3}\left[t_{dd1} + 2 t_{dd2} + (t_{dd2}-t_0)n_1\right]^2-(\Delta_1)^2\right], \\
J^{(0)}_{ab_3}&=\frac{4}{3U}\left[\frac{1}{3}\left[t_{dd1} + 2 t_{dd2} + (t_{dd2}-t_0)n_2\right]^2-(\Delta_1)^2\right].
\end{eqnarray}
\begin{eqnarray}
\nonumber
J^{(1z)}_{ab_1}&=\frac{8(\Delta_1)^2}{3U};\\
\nonumber
J^{(1z)}_{ab_2}&=\frac{8(\Delta_1)^2}{3U};\\
J^{(1z)}_{ab_1}&=\frac{8(\Delta_1)^2}{3U};
\end{eqnarray}
where we have taken the direction of the trigonal distortion is taken to be uniform.
\subsection{Second nearest neighbour}
These are shown in Fig. \ref{n3lattice}. These indirect hoppings are mediated by the Na atoms. In general, in presence of the trigonal distortion in the $[-1,-1,1]$ direction, the magnitude of the hopping amplitudes are also direction dependent. However, since the magnitudes themselves are expected to be small we shall neglect such directional dependence in the magnitudes. The result is summarized in Table \ref{tab_n3lattice}. 

\begin{table}[h!]
\centering
\subtable[NNN:$t_{am;a_1m'}/t_{am;a_4m'}$]{
\begin{tabular}{|c|c|c|c|}\hline
$ _{m'}\backslash ^m$ & $d_{xy}$ & $d_{yz}$ & $d_{zx}$ \\ \hline
$d_{xy}$ & -& $t_{2}+\Delta_2$ & - \\ \hline
$d_{yz}$ & $t_2-\Delta_2$ & - & - \\ \hline
$d_{zx}$ & - & - & -\\ \hline
\end{tabular}
}
\subtable[NNN:$t_{am;a_2m'}/t_{am;a_5m'}$]{
\begin{tabular}{|c|c|c|c|}\hline
$ _{m'}\backslash ^m$ & $d_{xy}$ & $d_{yz}$ & $d_{zx}$ \\ \hline
$d_{xy}$ & -& - & $t_2+\Delta_2$ \\ \hline
$d_{yz}$ & - & - & - \\ \hline
$d_{zx}$ & $t_2-\Delta_2$ & - & -\\ \hline
\end{tabular}
}
\subtable[NNN: $t_{am;a_3m'}/t_{am;a_6m'}$]{
\begin{tabular}{|c|c|c|c|}\hline
$ _{m'}\backslash ^m$ & $d_{xy}$ & $d_{yz}$ & $d_{zx}$ \\ \hline
$d_{xy}$ & - & - & -\\ \hline
$d_{yz}$ & - & - & $t_2+\Delta_2$ \\ \hline
$d_{zx}$ & - & $t_2-\Delta_2$ & -\\ \hline
\end{tabular}
}
\caption{Hopping paths for the second nearest neighbours}
\label{tab_n3lattice}
\end{table}

So for the $e'_g$ basis, we have
\begin{eqnarray}
\nonumber
T^{(02)}_{a,a_1}=T^{(02)}_{a,a_4} &=-\frac{t_2}{3}n_2, \ \ T^{(z2)}_{a,a_1}=T^{(z2)}_{a,a_4} =-\frac{\Delta_2}{\sqrt{3}}n_2\\
\nonumber
T^{(02)}_{a,a_2}=T^{(02)}_{a,a_5} &=-\frac{t_2}{3}n_1, \ \ T^{(z2)}_{a,a_2}=T^{(z2)}_{a,a_5} =-\frac{\Delta_2}{\sqrt{3}}n_1\\
T^{(02)}_{a,a_3}=T^{(02)}_{a,a_6} &=-\frac{t_2}{3}n_3, \ \ T^{(z2)}_{a,a_3}=T^{(z2)}_{a,a_6} =-\frac{\Delta_2}{\sqrt{3}}n_3
\end{eqnarray}
For example, tight binding fit of the DFT data uses only $t_2$ and finds $t_2\approx -0.075\ eV$ {\cite{2009_jin,2009_chkim}}. Therefore we have:
\begin{eqnarray}
J^{(2)}_{a,a_\alpha}=\frac{4}{3U}\left[\frac{(t_2)^2}{3}-(\Delta_2)^2\right],\ \ J^{(2z)}_{a,a_\alpha}=\frac{8(\Delta_2)^2}{3U}.
\end{eqnarray}
\subsection{Third nearest neighbour}
The third nearest neighbours are listed in Fig. \ref{n4lattice}. The hopping to the third nearest neighbour is mediated by the Na atoms. Again we shall neglect the directional dependence and take these to be in the magnitudes of the hoping amplitudes. The result is summarized in table \ref{tab_n4lattice}.
\begin{table}[h!]
\centering
\subtable[NNNN:$t_{am;b_1'm'}$]{
\begin{tabular}{|c|c|c|c|}\hline
$ _{m'}\backslash ^m$ & $d_{xy}$ & $d_{yz}$ & $d_{zx}$ \\ \hline
$d_{xy}$ & $t_3(b_1')$ & - & - \\ \hline
$d_{yz}$ & - & - & - \\ \hline
$d_{zx}$ & - & - & -\\ \hline
\end{tabular}
}
\subtable[NNNN:$t_{am;b_2'm'}$]{
\begin{tabular}{|c|c|c|c|}\hline
$ _{m'}\backslash ^m$ & $d_{xy}$ & $d_{yz}$ & $d_{zx}$ \\ \hline
$d_{xy}$ & -& - & - \\ \hline
$d_{yz}$ & - & $t_3(b_2')$ & - \\ \hline
$d_{zx}$ & - & - & -\\ \hline
\end{tabular}
}
\subtable[NNNN: $t_{am;b_3'm'}$]{
\begin{tabular}{|c|c|c|c|}\hline
$ _{m'}\backslash ^m$ & $d_{xy}$ & $d_{yz}$ & $d_{zx}$ \\ \hline
$d_{xy}$ & - & - & -\\ \hline
$d_{yz}$ & - & - & - \\ \hline
$d_{zx}$ & - & - & $t_3(b_3')$\\ \hline
\end{tabular}
}
\caption{The hoppings for the third nearest neighbours}
\label{tab_n4lattice}
\end{table}

 Tight-binding fit to the DFT results {\cite{2009_jin,2009_chkim}} indeed show that this hopping energy scale is of the order of
\begin{eqnarray}
t_3(b_\alpha')=t_n\approx-0.075\ eV
\end{eqnarray}
Therefore we have:
\begin{eqnarray}
T_{ab_\alpha'}^{(03)}=\frac{t_n}{3},\ \ T_{ab_1'}^{(z3)}=0;
\end{eqnarray}

or,
\begin{eqnarray}
J^{(3)}_{ij}=\frac{4[t_n]^2}{9U};\ \ J^{(3z)}_{ij}=0;
\end{eqnarray}

\section{Spin Wave Spectrum}
\label{appen_HP_boson}

To calculate the spin wave spectrum for the zig-zag state we use the usual Holstein-Primakoff method suited to collinear ordering which may alternate in direction. More precisely we introduce:
\begin{eqnarray}
S^z=S-a^\dagger a;\ \ S^+=\sqrt{2S}a;\ \ S^-=\sqrt{2S}a^\dagger
\end{eqnarray}
for one direction and 
\begin{eqnarray}
S^z=-S+a^\dagger a;\ \ S^+=\sqrt{2S}a^\dagger;\ \ S^-=\sqrt{2S}a
\end{eqnarray}
for the other direction. Since there are 4 sites per unit cell (refer Fig. 1(a) of the main text) the quadratic Hamiltonian is a $8\times 8$ matrix given by:
\begin{eqnarray}
H_Q=H_{cl}+H_{sp},
\end{eqnarray}
where $H_{cl}$ is the classical part dealt in the previous section. The spin wave Hamiltonian has the following form
\begin{eqnarray}
H_{sp}=\frac{S}{2}\sum{\bf k}\Psi_{\bf k}^\dagger\mathcal{H}_{\bf k}\Psi_{\bf k} + H_s
\end{eqnarray}
Here $\Psi_{\bf k}^\dagger=\left[a^\dagger_{{\bf k},1},a^\dagger_{{\bf k},2},a^\dagger_{{\bf k},3},a^\dagger_{{\bf k},4},a_{-{\bf k},1},a_{-{\bf k},2},a_{-{\bf k},3},a_{-{\bf k},4}\right]$ (the subscript $1,2,3,4$ refers to the four sites in the unit cell as shown in Fig. 1(a) of the main text) and
\begin{eqnarray}
H_s&=-\frac{S}{2}\left[(1-2x+5y)-(2\delta_2-\delta_1)\right]N_{Cell};\\ 
\mathcal{H}_{\bf k}&=\left[\begin{array}{cc}
A_{\bf k} & B_{\bf k}\\
B^\dagger_{\bf k} & A_{\bf k}
\end{array}\right]
\end{eqnarray}
where $N_{cell}$ is the number of unit cells and
\begin{eqnarray}
A_{\bf k}=\left[\begin{array}{cccc}
\chi_{\bf k} & 0 & 0 & \eta_{\bf k}\\
0 & \chi_{\bf k} & \phi_{\bf k} & 0 \\
0 & \phi^*_{\bf k} & \chi_{\bf k} & 0 \\
\eta^*_{\bf k} & 0 & 0 & \chi_{\bf k} \\
\end{array}\right];\ \ \ \ \
B_{\bf k}=\left[\begin{array}{cccc}
0 & \xi_{\bf k} & \rho_{\bf k} & 0\\
\xi^*_{\bf k} & 0 & 0 & \rho_{\bf k}\\
\rho^*_{\bf k} & 0 & 0 & \xi_{\bf k}\\
0 & \rho^*_{\bf k} & \xi_{\bf k} & 0\\
\end{array}
\right]
\end{eqnarray}
where,
\begin{eqnarray}
\chi_{\bf k}&=(2\delta_2-\delta_1)+(1-2x+5y+y\cos k_x);\\
\eta_{\bf k}&= x e^{\imath k_y}\left(1+e^{\imath k_x}\right);\\
\phi_{\bf k}&= x\left(1+e^{\imath k_x}\right);\\
\xi_{\bf k}&= (1+2y\cos{k_x}+ye^{-\imath k_y});\\
\rho_{\bf k}&= y(1+e^{\imath k_x})(1+e^{-\imath k_y}).
\end{eqnarray}
Now following usual methods we diagonalize
\begin{eqnarray}
\left[\begin{array}{cc}
A_{\bf k} & B_{\bf k}\\
-B^\dagger_{\bf k} & -A_{\bf k}\\
\end{array}\right]
\end{eqnarray}
to get the spin wave spectrum as plotted in Fig. 3(a) and 3(b) of the main text.
\section{Projection of Zeeman term in the $t_{2g}$ and $\left\{\vert e'_{1g}\downarrow\rangle,\vert e'_{2g}\uparrow \rangle\right\}$ subspaces.}
\label{appen_zeeman}
The Zeeman coupling term, when projected to the $t_{2g}$ space, gives
\begin{eqnarray}
H^{t_{2g}}_Z=\mu_B\left(-\vec l + 2\vec s\right)\cdot\vec B. 
\end{eqnarray}
Thus the magnetization after projection is given by:
\begin{eqnarray}
\vec M^{t_{2g}}=\mu_B\left(-\vec l+2\vec s\right)
\end{eqnarray}
This when projected to the subspace $\vert e'_1,\uparrow\rangle$ and $\vert e'_2,\downarrow\rangle$ gives (using the Block diagonal property of the orbital angular momentum as above):
\begin{eqnarray}
\tilde H_Z=4\mu_BS^z \hat n\cdot\vec B
\end{eqnarray}
where $\vec S$ (note that this is in upper case compared to the physical spin written lower case) is the emergent pseudo-spin-1/2 per site. 
This is the emergent degree of freedom at low energies. Clearly, the magnetic moment is then given by Eq. 7 of the main text.



\end{document}